\documentclass[11pt]{article}
\usepackage{moriond,epsfig}
\usepackage{wrapfig}

\def\be{\begin{equation}}
\def\ee{\end{equation}}
\def\bea{\begin{eqnarray}}
\def\eea{\end{eqnarray}}

\begin{document}
\vspace*{4cm}
\title{Study of QGP with probes associated with photon at RHIC-PHENIX} 

\author{Yoki Aramaki, for the PHENIX Collaboration}

\address{Center for Nuclear Study, Graduate School of Science,
  University of Tokyo, \\7-3-1 Hongo, Bunkyo, Tokyo 113-0033,
  Japan}  

\maketitle\abstracts{
When heavy ions with high energy collide, a hot and dense matter is
produced.  
As the matter expands, the matter undergoes cross-over phase
transition from partonic matter to hadronic matter. 
Jets are created by high $p_{T}$ partons in the early phase of the
collisions.
Leptons and photons can penetrate the matter without strong
interaction. 
For that reason, jets, leptons and photons are good probes to study
the partonic matter in the early phase of the collisions. 
We report about two probes associated with photon and their results
for PHENIX. 
}

\section{High $p_{T}$ neutral pions}
\subsection{Measurement of the initial density in the matter}
It has been observed in central Au+Au collisions at Relativistic
Heavy Ion Collider (RHIC) that the yield of neutral pion at high
transverse momentum ($p_{T} > 5$~GeV/$c$) is strongly suppressed
compared to the expectation from $p+p$ collisions scaled by the
number of binary collisions (Jet quenching).  
This suppression is regarded due to the energy loss of hard scattered
partons in the medium, which results in a decrease of the yield at a
given $p_{T}$.  
Many theoretical models have been proposed to understand the parton
energy loss mechanism. 
GLV method\cite{cite:GLV} as one of the calculations predicts that the
magnitude of energy loss is proportional to the square of the path
length.
Therefore measuring the path length dependence of energy loss should
improve our knowledge on energy loss constrain the parton energy loss
models.    

\subsection{Neutral pion production with respect to the reaction plane}
\begin{figure}[htbp]
  \centering
   \includegraphics[width=10.5cm]{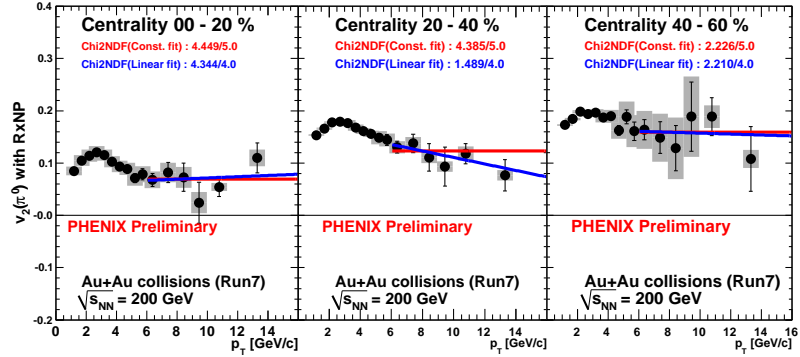}
    \caption{The anisotropy $v_{2}$ of neutral pion as a function of
      $p_{T}$ for each 20~$\%$ centrality class.
      Red and blue lines which are fitted to the data from 6~GeV/$c$
      show constant and linear fit function, respectively. }
    \label{Fig:v2fitting}
\end{figure}

To quantify the energy loss in the matter, we introduce the nuclear
modification factor ($R_{AA}$).
\begin{eqnarray}
R_{AA} = \frac{d^{2}N^{AA}/dp_{T}d\eta}{T_{AA}d^{2}\sigma^{NN}/dp_{T}d\eta},
\end{eqnarray}
where $N^{AA}$, $\sigma^{NN}$ and $T_{AA}$ are yields of neutral pion
in Au+Au collisions, the cross section of neutral pion ($p+p$) and the
nuclear overlap function calculated by Glauber model, respectively.  
The path length can be estimated by measuring the azimuthal angle 
from reaction plane and the impact parameter (centrality).   
Therefore the $R_{AA}$ for each azimuthal angle ($\Delta\phi_{i}$)
bin is measured to obtain the path length dependence of $\pi^{0}$
production.   
The $R_{AA}$($p_{T},\Delta\phi$) is derived by the following equations. 
\begin{eqnarray}
R_{AA}(p_{T},\Delta\phi_{i})&=&
R_{AA}(p_{T})\times
\frac{N(p_{T},\Delta\phi_{i})}{\Sigma_{\phi_{i}}N(p_{T},\Delta\phi_{i})}\\
N(p_{T},\Delta\phi_{i}) &\propto& 1+2v_{2}\cos(2\Delta\phi_{i}) 
\end{eqnarray}
The second harmonic term ($v_{2}$) means the azimuthal anisotropy. 
The anisotropy $v_{2}$ at low $p_{T}$ creates collective flow and the
flow is background for measuring the $R_{AA}(p_{T},\Delta\phi)$.
In order to reduce the effect of collective flow, we measure the
yields at higher $p_{T}$. 
Figure~\ref{Fig:v2fitting} shows the anisotropy $v_{2}$ of neutral
pion as a function of $p_{T}$ for each 20~$\%$ centrality class.
These values are non-zero for all centrality classes.
Additionally, assumed linear and constant functions are fitted to this
data and the fit results are shown in Table~\ref{table:FitResult}.
These results indicate that the anisotropy $v_{2}$ of neutral pion in most
central and peripheral collisions tends to be constant, while in
mid-central collisions it tends to decrease.   

\begin{table}[!htb]
  \begin{center}
    \begin{tabular}{|c|c|c|}
      \hline
      Centrality [$\%$] & Constant ($\chi^{2}$/NDF) & Linear ($\chi^{2}$/NDF)\\
      \hline
      00--20 & 4.45/5 & 4.34/4 \\
      20--40 & 4.39/5 & 1.49/4 \\
      40--60 & 2.23/5 & 2.21/4 \\
      \hline
    \end{tabular}
  \end{center}
    \caption{Fit results with the two functions for three centrality classes.}
  \label{table:FitResult}
\end{table}

\subsection{Comparison of $R_{AA}(p_{T},\Delta\phi)$}
\begin{figure}[htbp]
  \centering
  \includegraphics[width=12cm]{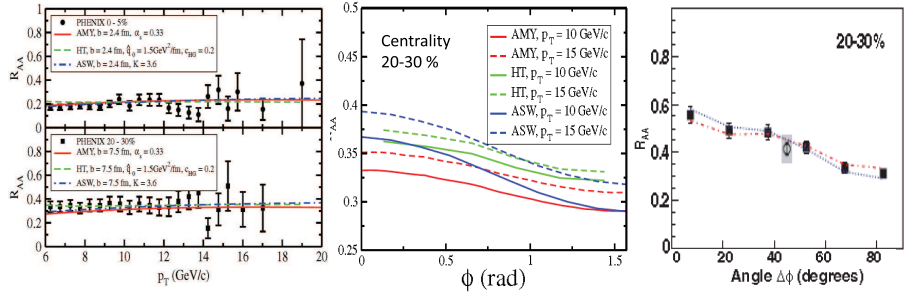}
  \caption{
    Left: $R_{AA}(p_{T})$ at centrality 0-5~$\%$ (top) and 20-30~$\%$
    (bottom) calculated in three models compared with the data. 
    Central: $R_{AA}(p_{T},\Delta\phi)$ at $p_{T} = 10$~GeV/$c$ (solid
    line) and $p_{T} = 15$~GeV/$c$ (dashed line) 
    for three models at centrality 20-30~$\%$.
    Right: $R_{AA}(p_{T},\Delta\phi)$ at $5 < p_{T} < 8$~GeV/$c$ at
    centrality $20-30\%$.
  }
  \label{Fig:RaaComp}
\end{figure}

Recently theoretical models (ASW\cite{cite:ASW}, HT\cite{cite:HT} and
AMY\cite{cite:AMY}) which involve the space-time evolution of the
matter have been proposed to investigate parton energy loss
mechanism\cite{cite:BassModelComp}.   
The left panel of Fig.~\ref{Fig:RaaComp} shows these models succeed in
reproducing the centrality dependence of $R_{AA}(p_{T})$.   
The central and right panels show the theoretical curves and the
measured data at the same centrality class, respectively.
These models are still unable to reproduce the measured
$R_{AA}(p_{T},\Delta\phi)$. 

\section{Low $p_{T}$ direct photon}
\subsection{Measurement of the initial temperature of the matter}
Thermally equilibrated matter is expected to radiate thermal photons,
whose kinematics ($p_{T}$ or yield) is sensitive to the temperature of
the matter. 
Thermal photons are predicted to be dominant in direct photons for low
$p_{T}$ ($1 < p_{T} < 3$~GeV/$c$) in Au+Au collisions.
On the other hand, a huge background is also produced by decay
photons at this $p_{T}$ region. 
Electron pairs via virtual photon have been measured by the PHENIX
experiment in order to improve the signal to background and associated
systematic errors\cite{cite:EnhanceDielectron}.

\subsection{Low mass $e^{+}e^{-}$ via internal conversion}
\begin{wrapfigure}{l}{8cm}
   \includegraphics[width=8cm]{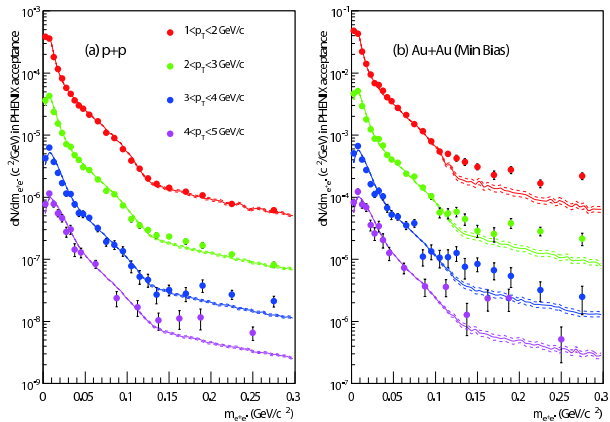}
    \caption{Left: $e^{+}e^{-}$ pair invariant mass distributions in
      (a) $p+p$ and (b) minimum bias Au+Au collisions, respectively.  
      The $p_{T}$ ranges are shown in the legend.
      Solid and dashed lines represent the expected values from the cocktail
      calculation and their systematic uncertainties, respectively.
    }
    \label{Fig:Dielectron}
\end{wrapfigure}

In general, any source of real photons can emit virtual photons and
the virtual photons can convert to low mass $e^{+}e^{-}$ pairs.
The relation of real photon production and $e^{+}e^{-}$ pair
production can be derived by Knoll-Wada formula\cite{cite:Knoll-Wada}
based on quantum electrodynamics (QED).   
Figure.~\ref{Fig:Dielectron} shows mass spectra of $e^{+}e^{-}$ pairs
in $p+p$ and Au+Au collisions for several $p_{T}$ ranges. 
These mass spectra are compared to expected yields from a cocktail of
the hadron decay calculation.
This cocktail calculation is based on meson production as measured by
PHENIX.  
The $p+p$ data are consistent with the cocktail for low $p_{T}$
region, while they have a small excess for high $p_{T}$ region. 
The Au+Au data are good agreement with the cocktail for $M_{ee} <
50$~MeV/$c^{2}$ and the excess of the yields have much larger than
$p+p$ for all $p_{T}$ bins.

\begin{figure}[htbp]
  \centering
    \includegraphics[width=11cm]{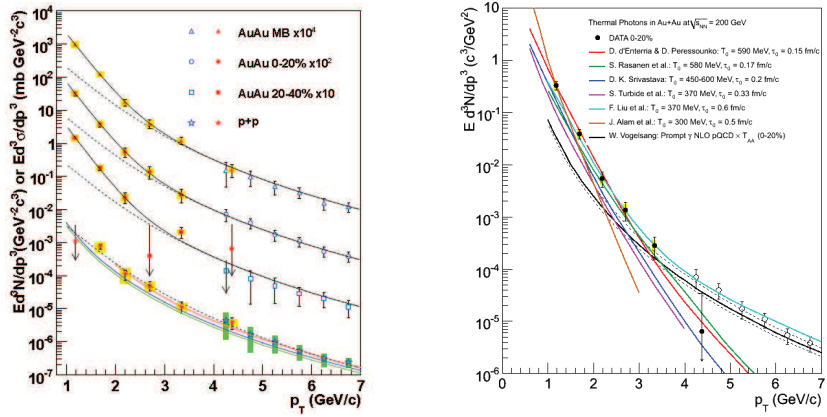}
    \caption{The left panel shows invariant cross section ($p+p$) and
      invariant yield (Au+Au) of direct photon as a function of $p_{T}$.
      The right panel shows invariant cross section for Au+Au
      collisions (centrality $0-20\%$) and several calculations. } 
    \label{Fig:DirectPhoton}
\end{figure}

Red, blue and green lines show next-to-leading-order perturbative
quantum chromodynamics (NLO pQCD) calculations which include
theoretical scales for $\mu = 0.5~p_{T}$, $p_{T}$ and $2~p_{T}$,
respectively.  
The $p+p$ data are consistent with the expectation from NLO pQCD
calculations, while the clear enhancement for the Au+Au data is shown
at $p_{T} < 3.5$~GeV/$c$.
The direct photon invariant yield is extracted from the fraction of
the direct photon yield and the inclusive photon yield.  
Figure.~\ref{Fig:DirectPhoton} shows the invariant cross section of
photon in $p+p$ and as a function of $p_{T}$ in $p+p$ and the
invariant yield for several centrality classes in Au+Au collisions. 
The filled and open points are from virtual photon analysis and from
the Ref.~\cite{cite:AuAuGamma,cite:ppGamma}, respectively.  
The three curves on the $p+p$ data represent NLO pQCD calculations and
the dashed lines show modified power-law fit to the $p+p$ data, scaled
$T_{AA}$. 
The black lines are exponential plus the $T_{AA}$-scaled $p+p$ fit.
The NLO pQCD calculation is consistent with the $p+p$ data within
theoretical uncertainties for $p_{T} > 2$~GeV/$c$.
As shown in Fig.~\ref{Fig:DirectPhoton}, the $p+p$ data can be
well described by a modified power-law fit function
($A_{pp}(1+p_{T}^{2}/b)^{-n}$).  
Furthermore, we fit an exponential plus the $T_{AA}$-scaled modified
power law function ($Ae^{-p_{T}/T}+T_{AA}\times
A_{p+p}(1+p_{T}^{2}/b)^{-n}$) to the direct photon invariant yield in
Au+Au collisions. 
Here $T_{AA}$ is the nuclear overlap function and the free parameters
are $A$ and the inverse slope $T$. 
This inverse slope $T$ is closely related to the temperature in the
matter.
The obtained $T$ from the fit is $221 \pm 23 \pm 18$~MeV for central
Au+Au collisions.   
This value is beyond the critical temperature ($T_{c}$) expected from
lattice QCD ($T_{c}\approx$170~MeV).  
The right panel of Fig.~\ref{Fig:DirectPhoton} shows
the comparison of invariant cross section for Au+Au central collisions
and the direct photon spectra obtained by several hydrodynamical
calculations\cite{cite:HydroComp}.  
Hydrodynamical calculations assume initial temperature ranging from
$T_{init} = 300$~MeV at thermalization time $\tau_{0} = 0.6$~fm/$c$ to
$T_{init} = 600$~MeV at $\tau_{0} = 0.15$~fm/$c$. 
These models can reproduce the central Au+Au collisions (centrality
$0-20\%$) data within a factor of two.

\section{Summary}
The probes associated with photon are a powerful tool to study the
property of hot and dense matter.
For studying parton energy loss mechanism, path length dependence of
the energy loss for high $p_{T}$ hadrons is expected to strongly
constrain the parton energy loss calculations. 
The azimuthal anisotropy $v_{2}$ of neutral pion is measured for three
centrality classes and their values at high $p_{T}$ are finite.  
Theoretical calculations succeed to reproduce the centrality
dependence of $R_{AA}(p_{T})$, however azimuthal angle dependence of
$R_{AA}$ can not still be reproduced. 
For measuring the initial temperature of the matter, $e^{+}e^{-}$
pairs through internal conversion are measured at $1 < p_{T} <
5$~GeV/$c$.
The enhancement is clearly observed in Au+Au minimum bias events for
$1 < p_{T} < 5$~GeV/$c$.
The $T_{AA}$-scaled $p+p$ fit function is fitted to the direct photon
spectrum for the central Au+Au collisions.
The obtained inverse slope from the fit is $T = 221 \pm 23(stat.) \pm
18(sys.)$.  
This is beyond the temperature of a phase transition which lattice QCD
predicts.

\end{document}